# A New Method For A Sensitive Deuteron EDM Experiment

Y.K. Semertzidis[4], M. Aoki[13], M. Auzinsh[9], V. Balakin[2], A. Bazhan[2], G.W. Bennett[4], R.M. Carey[3], P. Cushman[10], P.T. Debevec[7], A. Dudnikov[2], F.J.M. Farley[15], D.W. Hertzog[7], M. Iwasaki[14], K. Jungmann[5], D. Kawall[15], B. Khazin[2], I.B. Khriplovich[2], B. Kirk[4], Y. Kuno[13], D.M. Lazarus[4], L.B. Leipuner[4], V. Logashenko[2,3], K.R. Lynch[3], W.J. Marciano[4], R. McNabb[10], W. Meng[4], J.P. Miller[3], W.M. Morse[4], C.J.G. Onderwater[7], Y.F. Orlov[6], C.S. Ozben[7], R. Prigl[4], S. Rescia[4], B.L. Roberts[3], N. Shafer-Ray[11], A. Silenko[1], E.J. Stephenson[8], K. Yoshimura[12]

EDM Collaboration

*1. Belarusian State University, Belarus; 2. BINP, Novosibirsk; 3. Boston University; 4. Brookhaven National Laboratory; 5. Kernfysisch Versneller Instituut, Groningen; 6. Newman Laboratory, Cornell University, Ithaca; 7. University of Illinois, at Urbana-Champaign; 8. Indiana University; 9. University of Latvia; 10. University of Minnesota; 11. University of Oklahoma; 12. KEK, Japan; 13. Osaka University, Japan; 14. RIKEN, Japan; 15. Department of Physics, Yale University, New Haven.*

**Abstract.** In this paper a new method is presented for particles in storage rings which could reach a statistical sensitivity of $10^{-27}$ e·cm for the deuteron EDM. This implies an improvement of two orders of magnitude over the present best limits on the T-odd nuclear forces $\xi$ parameter.

## INTRODUCTION

Sensitive searches for electric dipole moments (EDM) of fundamental particles are excellent probes of physics beyond the standard model (SM) [1]. The reason for this is because physics beyond the SM allows for values that are well within reach of present technology while the backgrounds from the SM predictions are many orders of magnitude below them.

In this paper a new method is presented for particles in storage rings which could reach a statistical sensitivity of $10^{-27}$ e·cm for the deuteron EDM. The method uses a radial electric field to cancel the g-2 precession of particles in order to maximize the EDM signal strength by several orders of magnitude [2,3]. The same technique is used to improve upon the muon EDM sensitivity level; see contribution by J. Miller elsewhere in these proceedings.

There are three candidate laboratories that the collaboration is considering at this point for the deuteron EDM experiment: Brookhaven National Laboratory (BNL),

Indiana University Cyclotron Facility (IUCF) and Kernfysisch Vernsneller Instituut (KVI) at Groningen of The Netherlands. In the numerical examples of this paper we will assume parameters which are relevant at the BNL site. The systematic errors are still under consideration.

## SPIN PRECESSION

The spin *s* of a particle with a magnetic moment $\mu$ and an electric dipole moment *d*, located in magnetic and electric fields *B* and *E*, precesses according to

$$\frac{d\vec{s}}{dt} = \vec{\mu} \times \vec{B} + \vec{d} \times \vec{E} \qquad (1)$$

In a storage ring with only B-field present and *d*=0, the spin precession rate relative to the momentum vector is given by

$$\vec{\omega}_a = \frac{e}{m} a \vec{B} \qquad (2)$$

with *a=(g-2)/2*, where *g* is the gyromagnetic ratio. The spin precession plane is in the same plane as the momentum precession. If *d*≠0 the above Equation 2 becomes

$$\vec{\omega}_T = \frac{e}{m} a \vec{B} + \frac{1}{\hbar} \vec{d} \times (\vec{u} \times \vec{B}) \qquad (3)$$

for spin 1 particles, where $(\vec{u} \times \vec{B})$ is the motional electric field, i.e. the E-field the particle feels in its own rest frame, $\vec{u} = \vec{\beta}/c$ the particle's velocity. The effect of the EDM component is to tilt the spin precession plane relative to the momentum plane at *every* azimuthal location. For a small value of *d* and a relatively large value of *a* the tilt is very small, e.g. for the deuteron, a=-0.143, and assuming a value for $d = 10^{-27}$ e·cm and $\beta$=0.7, then the tilt angle is $5 \times 10^{-13}$ rad which is very small. If the g-2 precession term was zero, the only precession would be due to the EDM and it would be entirely out-of-plane. Moreover, the spin phase angle would accumulate with time, proportional to the EDM. One way to eliminate the g-2 term is to apply an electric field equal to

$$E = \frac{aBc\beta}{1-(1+a)\beta} \approx aBc\beta\gamma^2 (1 + a\beta^2\gamma^2) \qquad (4)$$

where *c* is the speed of light and $\gamma$ the relativistic Lorentz factor [2,3]. If the g-2 cancellation is achieved at the $10^{-7}$ level then the tilt angle becomes $5 \times 10^{-6}$ rad, i.e. seven orders of magnitude higher than without the cancellation.

# EXPERIMENTAL PARAMETER VALUES

## EDM Signal Strength

In the absence of the g-2 precession, Equation (3) can be re-written as

$$\omega_d = \frac{1}{\hbar} duB \approx \frac{d}{\hbar a \gamma^2 (1 + a\beta^2\gamma^2)} E \qquad (5)$$

i.e. the EDM signal is proportional to the electric field strength $E$, inversely proportional to the anomalous magnetic moment $a$, and approximately inversely proportional to $\gamma^2$. Assuming a 5 cm plate separation, then an electric field of 4 MV/m is reasonable. For $d = 10^{-27}$ e·cm, and $\beta = 0.7$ the EDM signal becomes $\omega_d = 2.5 \times 10^{-7}$ rad/s.

## Storage Ring Parameters

At BNL one possible location for the deuteron EDM experiment is the AGS tunnel. Assuming 80% ring coverage with magnetic and electric fields, an electric field of 4MV/m and $\beta = 0.7$, $\gamma = 1.4$ then the required magnetic field is $B \approx 0.8$ KG, a rather low field. The radius of the tunnel is 127 m and there is space for another ring in the tunnel. The deuteron momentum is 2 GeV/c and its kinetic energy 0.9 GeV.

## Signal Detection

The out of plane spin precession can be detected by scattering the deuterons off a proton or carbon target and looking for a left-right asymmetry versus time. We will assume a carbon target here. The deuteron cross section on carbon targets has been measured and analyzed [5,6,7,8] at 1.69 GeV/c or 650 MeV kinetic energy of deuteron, close to the parameters we are considering here.

The analyzing power for deuterons on carbon has been measured with the POMME polarimeter in the semi-inclusive [9] and inclusive [10] mode in the energy range of 0.175 to 1.6 GeV using the polarized deuteron beam of the Laboratoire National Saturn in France. In the interesting range $3^0$-$13^0$ the elastic cross section is estimated from ref. [5] to be (40±10) mb or about 30% of the total elastic.

### *Expected Rates in the AGS Tunnel and Estimated Statistical Error*

In our case we want to use a thin carbon target and detect in-time coincidence between the deuteron and recoiling carbon as a function of storage time. The asymmetry is expected to be at least as much as measured by the semi-inclusive method [9].

Preliminary M.C. studies showed that the useful elastic scattering rate ratio is about $f \approx 3\times10^{-4}$. If we have 100 carbon target stations distributed around the ring, the counting rate on the DAQ will be reduced by the same factor. The estimated polarized deuteron beam intensity is $12\times10^{11}$ per cycle [11], and making the reasonable assumption that the distributed DAQ can take the rate, the useful rate is then $\approx 3\times10^{8}$ per cycle with an average asymmetry of at least 0.35. The beam polarization is expected to be 85% of the maximum 2/3 for a "pure" vector polarized beam, i.e. 0.56 [11].

The deuteron beam will de-polarize as a function of time due to E, B-field multipoles, betatron oscillations, momentum dispersion, etc. At this point we believe we can achieve a polarization lifetime of 10 s. One can then optimize the beam lifetime (the beam is lost as a result of multiple scattering in the target), and the beam measuring time per cycle for best statistical sensitivity. The statistical error on the deuteron EDM when the above parameters are optimized is given by

$$\sigma_d = 4.6 \frac{\hbar a \gamma^2 (1 + a\beta^2\gamma^2)}{E[1 + a\gamma^2(1 + a\beta^2\gamma^2)] AP\sqrt{NfT_{tot}\tau}} \quad (6)$$

where $\tau$ is the beam lifetime, $E$ is the radial electric field value, $A$ is the polarimeter asymmetry, $P$ is the deuteron beam polarization, $\hbar = 6.58\times10^{-22}$ MeV s, $a = -0.143$ the deuteron anomalous magnetic moment, $\gamma^2 \approx 2$, $N$ is the average number of stored particles per cycle, $f$ the fraction of useful events, and $T_{tot}$ the total running time of the experiment. Assuming $\tau = 5$ s, $E_R = 4$ MV/m with 80% coverage of the ring with the E-field, and a total $10^7$ s running time, then $\sigma_d \approx 1.1\times10^{-27}$ e·cm statistical error. Equation (6) is valid for the deuteron EDM experiment when the polarization lifetime is 10 s, the beam lifetime is 5 s and the measuring time is 10 s.

## Systematic Errors

The main systematic error is due to an out of plane electric field component [2,3] and the background spin precession rate is given by

$$\omega_B = \frac{geE}{2m\beta c\gamma^2}\theta_E \quad (7)$$

with $g$ the gyromagnetic ratio and $\theta_E$ the misalignment angle. We are planning to inject the particles into the ring [3] clockwise (CW) and counter clockwise (CCW) to cancel the effect. We are also studying the possibility to store polarized protons in the same ring to study and optimize the electric field directional stability since their sensitivity to the misalignment angle is more than 10 times greater than that of the deuterons. Using the same electric field with opposite polarity the protons would have a momentum of about 0.5 GeV/c and the required magnetic field would be about 50 Gauss. The intensity of the polarized protons is expected to be similar to the intensity of the polarized deuterons [11] and the results would be easier to interpret since it is a spin ½ particle and has no tensor polarization components.

# THEORETICAL IMPLICATIONS

Improving the deuteron EDM sensitivity to the level of low $10^{-27}$ e·cm, would be a very interesting experiment. The parameter $\xi$ of the T-odd nuclear forces is related [12] to the electric dipole moment of the deuteron according to the equation $d = 2 \times 10^{-22} \xi$ e·cm. The best limit on $\xi$ of $0.5 \times 10^{-3}$ currently comes form the $^{199}$Hg EDM experiment [13]. A deuteron EDM experiment at the intended sensitivity level would be an improvement of two orders of magnitude over that limit. Also, since the deuteron consists of a proton and a neutron, at that level it would be improving upon the neutron EDM by a factor of 60 to 100 and more than four orders of magnitude upon the proton EDM.